\documentclass[aps,superscriptaddress,twocolumn,showpacs]{revtex4-1}
\pdfoutput=1
\usepackage{color}
\usepackage{amsmath, amsthm, amsfonts}    
\usepackage{amssymb}
\usepackage{mathtools}
\usepackage{mathptmx} 
\usepackage[table]{xcolor}
\usepackage{dsfont}
\usepackage{enumitem} 
\usepackage{appendix}
\usepackage{braket}
\usepackage[skins,theorems]{tcolorbox}
\tcbset{highlight math style={enhanced,
  colframe=red,colback=white,arc=0pt,boxrule=1pt}}
\usepackage{cancel}
\usepackage{empheq}
\usepackage{lipsum}
\usepackage{graphicx}
\usepackage{tikz}
\usepackage{makecell}
\usepackage{siunitx}

\normalsize
\usepackage{booktabs}
\providecommand{\Eq}[1]{Eq.~(\ref{#1})}
\providecommand{\alphaem}{\alpha_{\text{em}}}
\providecommand{\alphaee}{\alpha_{\text{ee}}}
\providecommand{\alphamm}{\alpha_{\text{mm}}}
\providecommand{\ii}{\mathrm{i}}

\usepackage[unicode=true,pdfusetitle,
 bookmarks=true,bookmarksnumbered=false,bookmarksopen=false,
 breaklinks=true,pdfborder={0 0 0},backref=false,colorlinks=true,citecolor=blue]{hyperref}
\usepackage{graphicx}   
\usepackage[caption=false]{subfig}       
\usepackage{bm}            
\usepackage[normalem]{ulem}  

\newlength\imagewidth
\newlength\imagescale
\def\be{\begin{eqnarray}}
\def\ee{\end{eqnarray}}
\def\r{{\bf r}}

\def\v{{\bf v}}

\def\E{{\bf E}}
\def\H{{\bf H}}

\def\D{{\bf D}}
\def\p{{\bf p}}
\def\m{{\bf m}}

\def\im{{\rm i}}
\newcommand*{\til}[1]{{\overline{#1}}}

\definecolor{JOT-color}{named}{blue}
\definecolor{CSF-color}{named}{orange}

\begin{document}

\title{A Novel Chiroptical Spectroscopy Technique}

\author{Jorge Olmos-Trigo}
\email{jolmostrigo@gmail.com}
\affiliation{Faculty of Optics and Optometry, Universidad Complutense de Madrid, Madrid
28037, Spain.}

\author{Cristina Sanz-Fernández}
\affiliation{Centro de F\'isica de Materiales (CFM-MPC), Centro Mixto CSIC-UPV/EHU, Manuel de Lardizabal 4, 20018 San Sebasti\'an, Spain.}

\author{Ivan Fernandez-Corbaton}
\affiliation{
Institute of Nanotechnology, Karlsruhe Institute of Technology, 76021 Karlsruhe, Germany.}

\begin{abstract}
Chiral objects typically exhibit a different extinction for the two circular polarizations of light. Researchers often detect the chirality of objects by measuring this extinction difference employing Circular Dichroism (CD) spectroscopy. In this Letter, we present a new spectroscopy technique for detecting the chirality of spherical objects based on measuring the Stokes parameters at any non-forward angle. The chirality measure we introduce effectively eliminates achiral background noise and is independent of both the object's concentration and the optical path length. Notably, when a solution contains both enantiomers of a chiral object, our method can discern which enantiomer predominates. Furthermore, we demonstrate that the technique is robust and verifiable in-situ by measuring the Stokes vector at two different non-forward angles of choice.
\end{abstract}

\maketitle

Chirality refers to the geometric property of objects that cannot be superimposed onto their mirror images. In nature, chirality is ubiquitous; numerous organic molecules, such as glucose, and most biological amino acids, are chiral. 
In the pharmaceutical industry, chiral specificity is crucial as opposite enantiomers, mirror pairs of chiral molecules, can have vastly different effects on biological systems depending on their handedness~\cite{lenz1962thalidomide}. Enantiomer pairs share the same atomic composition and are indistinguishable when measuring their scalar molecular properties. It is through interactions with other chiral entities that their chirality can be revealed.

In nanophotonics, the electromagnetic helicity is the most commonly used chiral entity to unveil the chirality of objects~\cite{fernandez2013electromagnetic}. 
Circular Dichroism (CD) spectroscopy stands as the most utilized technique to unveil molecular chirality using helicity~\cite{barron2009molecular, adhikari2022optically}. In a conventional CD setup, the molecular solution is sequentially illuminated with fields of opposite helicities, and the total transmitted power is recorded for each case. The CD signal is then computed by taking the difference between these two power measurements in transmission, the forward direction of the light-scattering system.

Despite its widespread commercial use, CD exhibits drawbacks. For instance, if CD = 0 for a given frequency of the incident electromagnetic field, it is indeterminate whether the molecular solution is chiral or not. This ambiguity arises because CD provides only partial information about the chirality of the molecule~\cite{barron2009molecular}. 
To overcome this limitation and gain more insights into the chiral nature of the molecular solution, researchers typically measure optical rotation (OR), which 
quantifies the change of the polarization direction when linearly polarized light propagates through a chiral medium. 
However, both CD and OR are measured in the forward direction, which presents a challenge: the presence of a large achiral background signal that hinders the accuracy and reliability of measurements. Moreover, CD and OR depend on molecular concentration and optical path length, which makes the signature of CD and OR measurements of a given chiral object non-universal.

\begin{figure}[t!]
\centering\includegraphics[width=0.975\linewidth]{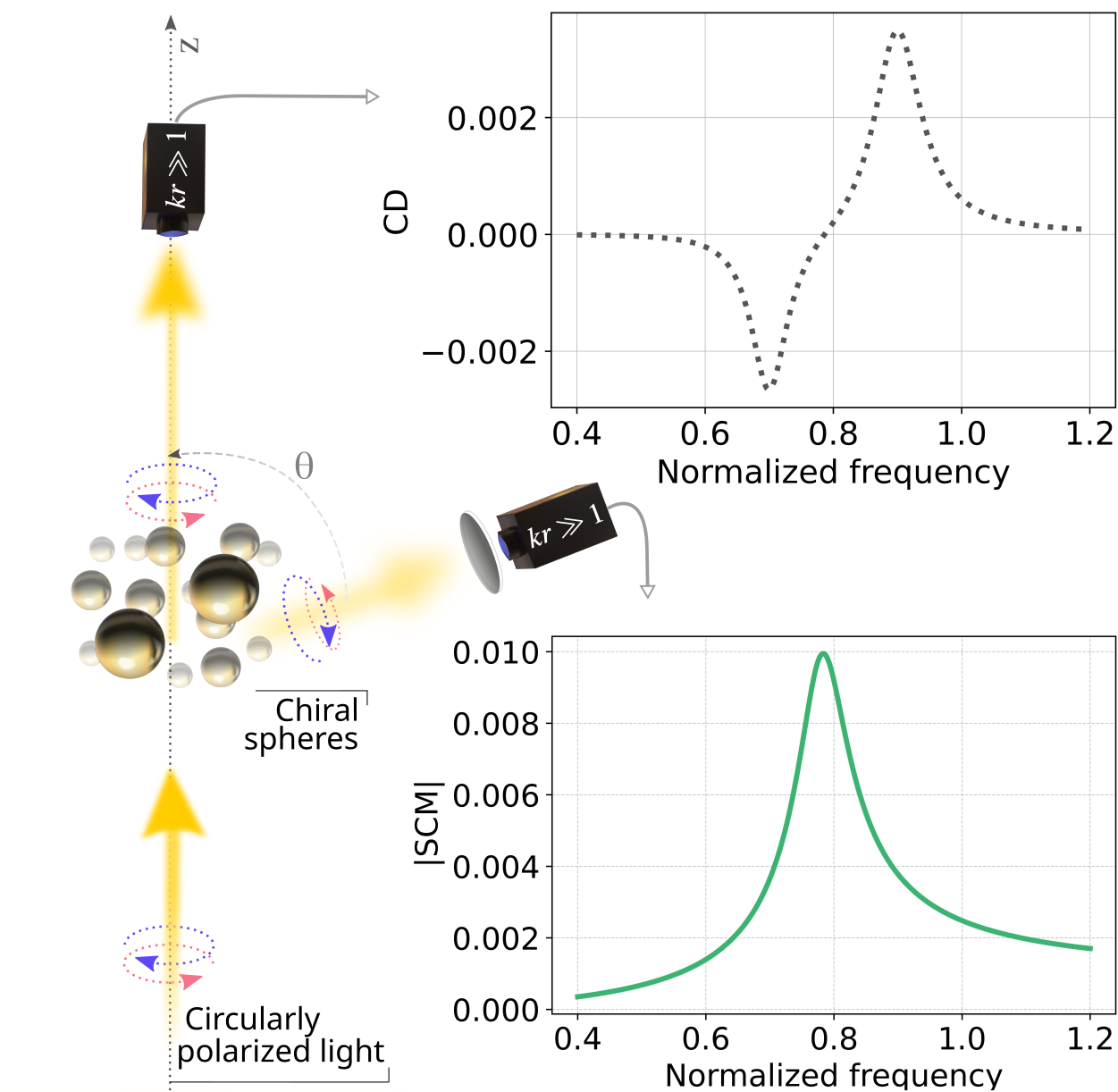}
    \caption{Chiral sphere sequentially illuminated by electromagnetic fields of opposite helicities. Two chiral observables are recorded: Circular Dichroism (CD) in transmission and the Stokes Chirality Measure (SCM) using a Stokes vector measurement.} 
    \label{fig:enter-label}
\end{figure}

In this Letter, we introduce a novel spectroscopy technique for detecting the chirality of spherical objects based on measuring the Stokes parameters at non-forward angles. The chirality measure that we introduce is universal, and in particular, avoids the achiral background and is independent of the object's concentration and the optical path length. Additionally, when the solution presents both enantiomers of a chiral object, our polarimetry-based method can discern which enantiomer predominates in the solution.

Importantly, we demonstrate that our novel technique is robust, and its accuracy can be experimentally verified in situ by measuring the Stokes parameters at two different non-forward angles of choice. Our findings pave the way to characterize the chirality of objects beyond CD and OR, the current state-of-the-art chiroptical spectroscopy techniques.

\emph{The setting.}---
The interaction between the electromagnetic field and a small object, such as most chiral molecules, 
can be treated in the linear dipolar approximation. This approximation
relates a 6x6 polarizability tensor with the incident
electric and magnetic fields $\{\E_{\rm{inc}}(\r), Z\H_{\rm{inc}}(\r) \} \in \mathbb{C}^3
$ at the position
of the object, $\r = \r_0$, with the induced electric and magnetic 
dipolar moments, denoted as $\p(\r_0)$ and $\m(\r_0)$, respectively: 
\begin{align}\label{Rel}
\begin{pmatrix}
\p (\r_0) / \epsilon \\
Z \m (\r_0)
\end{pmatrix}
= 
\begin{pmatrix}
\til{\alpha}_{\rm{ee}} & \til{\alpha}_{\rm{em}}  \\
\til{\alpha}_{\rm{me}} & \til{\alpha}_{\rm{mm}}  
\end{pmatrix}
\begin{pmatrix}
\E_{\rm{inc}}(\r_0)\\
Z \H_{\rm{inc}}(\r_0)
\end{pmatrix}.
\end{align}
Here, all $\til{\alpha}_{\rm{ij}}$ are 3x3 complex-valued tensors, and $\epsilon$ and $Z = \sqrt{\mu / \epsilon}$ denote the electric permittivity and impedance of the surrounding medium, respectively. 
Note that the electric $\til{\alpha}_{\rm{ee}}$, magnetic $\til{\alpha}_{\rm{mm}}$, and chiral $\til{\alpha}_{\rm{em}, \rm{me}}$ polarizabilities have dimensions of volume in our formulation. A harmonic time dependence $\exp(-\ii\omega t)$ is assumed and suppressed from the notation. The $\til{\alpha}_{\rm{ab}}$ tensors are frequency dependent, as are some other quantities, but we only write such dependence later. 

At this point, it is very often assumed that the rotationally averaged response of a dipolar object can be approximated by the response of an isotropic dipolar object~\cite{tang2010optical, Sersic2011, Scott2020}. It is important to realize that such an approximation is typically good if and only if such a response is evaluated in the forward direction. At non-forward angles, this approximation generally does not hold. 
An exception to this rule is a spherical nanoparticle, which is the system we consider in the following derivation. We should also mention that the spherical chiral model, derived by Bohren in 1974~\cite{bohren1974light}, is the most widely used framework for studying chiral light–matter interactions~\cite {yokota2001scattering,canaguier2013mechanical,hayat2015lateral, li2019optical, klimov2014eigen, wang2014lateral, kamandi2017enantiospecific,sifat2022force,ali2023enantioselective, kim2022molecular, sadrara2023large, shi2020chirality,mohammadi2023nanophotonic, martinez2024longitudinal, mohammadi2024nanophotonic, golat2024optical, tetour2024preparations}. 
Given all these considerations, we can express Eq.~\eqref{Rel} as
\begin{align}\label{Rel_1}
\begin{pmatrix}
\p(\r_0) / \epsilon \\
Z \m(\r_0)
\end{pmatrix}
= 
\begin{pmatrix}
{\alpha}_{\rm{ee}}\til{I} & {\alpha}_{\rm{em}}\til{I}  \\
-{\alpha}_{\rm{em}}\til{I} & {\alpha}_{\rm{mm}}\til{I}
\end{pmatrix}
\begin{pmatrix}
\E_{\rm{inc}}(\r_0)\\
Z \H_{\rm{inc}}(\r_0)
\end{pmatrix},
\end{align}
where $\til{I}$ is the 3x3 identity matrix. For scalars, as is the case, we have used that $\alpha_{\rm{em}} = -\alpha_{\rm{me}}$ because of reciprocity~\cite{onsager1936electric}.
Upon interaction with the incident field, dipoles typically re-radiate. 
The field scattered by an electric and magnetic dipole can be written in the radiation (far-field) zone as~\cite{jackson1999electrodynamics}
\begin{equation} \label{e_fields}
\E_{\rm{sca}}(\r) = k^2\frac{e^{\im kr}}{4 \pi r} \left[ \left(\hat{\mathbf{e}}_{\rm{r}} \times \frac{\p(\r_0)}{\epsilon} \right) \times \hat{\mathbf{e}}_{\rm{r}} -  \left(\hat{\mathbf{e}}_{\rm{r}} \times Z \m(\r_0) \right) \right].
\end{equation}
Here $\hat{\mathbf{e}}_{\rm{r}}$ denotes the radial unit vector, $k$ is the wavenumber in the surrounding medium, and  $r = |\r - \r_0|$ denotes the observation distance to the center of the dipolar object.
In general, we need to consider the summation of Eq.~\eqref{e_fields} for the total number of objects (hereafter denoted by N) dispersed in the solution. However, in the far field, any quadratic form of the scattered field results from an incoherent superposition of the individual contributions of the objects if these are randomly dispersed and do not interact with each other (see pages 461-462 of Ref~\cite{jackson1999electrodynamics}). 
This latter setting assumes that the field incident on any given scattering object is only that of the incident illumination, neglecting any contribution from scattered fields from the other objects. Such approximation is ubiquitous in the treatment of typical solutions of electromagnetically small objects. 
Therefore, to calculate the Stokes parameters, which are quadratic forms of the scattered field, we can simplify our analysis to consider only the field scattered by a single dipolar object located at $\mathbf{r}_0 = 0$.

We now introduce the Stokes vector and assume $\mathbf{r}_0 = 0$ in the induced dipoles and in the incident field.  

\emph{The Stokes Vector measurement.}---The Stokes vector $\mathbf{S} = [s_0, s_1, s_2,  s_3]$ unambiguously describes the polarization state in the far field ~\cite{stokes1851composition}. The components of the Stokes vector, often referred to as the Stokes parameters~\cite{crichton2000measurable, olmos2024spheres}, can be measured using conventional optical components such as a photodetector and waveplates~\cite{hinamoto2020colloidal}.
Following Crichton and Marston's notation (see Eqs.~(9a-d) of Ref~\cite{crichton2000measurable}), the Stokes parameters read as
\begin{alignat}{2} 
s_0 &= |E_\theta|^2 + |E_\varphi|^2, & \qquad
\label{s_01}
s_1 &= |E_\theta|^2 - |E_\varphi|^2, \\
s_2 &= -2\Re \{E_\theta E^*_\varphi \}, & \qquad
s_3 &= 2\Im \{E_\theta E^*_\varphi \}.
\label{s_23}
\end{alignat}
Here, $s_0$ is the total scattered intensity, $s_1$ is the degree of linear polarization, $s_2$ is the degree of linear polarization at $45^\circ$ degrees, and $s_3$ denotes the degree of circular polarization. Note that $\theta$ denotes the scattering and azimuth angles in spherical coordinates. In this regard, we use the same angle conventions as in Ref~\cite{crichton2000measurable}.

At this point, we have all the ingredients to calculate the polarizabilities as a function of the Stokes parameters given in Eqs.~\eqref{s_01}-\eqref{s_23}. We will assume that the solution is subsequently illuminated by two incident plane waves that differ only by their helicity. That is, by their circular polarization handedness $\sigma=\pm1$: $Z \H_{\rm{inc}}(\r) = -\im \sigma \E_{\rm{inc}}(\r)$~\cite{fernandez2013electromagnetic}, as schematically depicted in Figure.~\ref{fig:enter-label}. 
Expanding Eqs.~\eqref{s_01}-\eqref{s_23} using Eqs.~\eqref{Rel_1}-\eqref{e_fields}  we can get an equation ({see Supporting Information S1}) of the form $\mathbf{J}^{\sigma} =   \til{U}^\sigma (kr, \theta)  {{\mathbf{S}}^\sigma(kr, \theta)}$, where

\begin{widetext}
  \begin{equation} \label{compact} N
 \underbrace{\begin{pmatrix}
 |\gamma^\sigma|^2 \\
 |\beta^\sigma|^2\\ \Re \{\gamma^\sigma (\beta^\sigma)^* \}\\
    \Im \{\gamma^\sigma (\beta^\sigma)^* \}
    \end{pmatrix}}_{ \let\scriptstyle\textstyle\substack{\mathbf{J}^\sigma}} = \underbrace{\frac{32 \pi ^2 (kr)^2}{k^6 |E_0|^2 }  \frac{\csc^4 \theta}{2}
    \begin{pmatrix}
    1 + \cos^2 \theta & - \sin^2 \theta& 0 & 2 \sigma \cos \theta \\
    1 + \cos^2 \theta  & \sin^2 \theta& 0 &  2 \sigma \cos \theta  \\
    -2 \cos \theta & 0 & 0 & -\sigma \left(1 + \cos^2 \theta \right)\\
    0 & 0 & \sigma \sin^2 \theta & 0  
\end{pmatrix}}_{\let\scriptstyle\textstyle\substack{{\til{U}^\sigma (kr, \theta)}}}
\underbrace{
 \begin{pmatrix}
    {s}^\sigma_0 (kr, \theta)\\
    {s}^\sigma_1 (kr, \theta)\\ {s}^\sigma_2 (kr, \theta)\\
    {s}^\sigma_3 (kr, \theta)
\end{pmatrix}}_{\let\scriptstyle\textstyle\substack{{{\mathbf{S}}^\sigma(kr, \theta)}}}.
\end{equation}
Here we have defined 
\begin{align} \label{gamma}
\gamma^\sigma =\alpha_{\rm{ee}} - \im \sigma \alpha_{\rm{em}}, &&        \beta^\sigma =\alpha_{\rm{mm}} - \im \sigma \alpha_{\rm{em}}, 
\end{align}
\end{widetext}

Equation~\eqref{compact} is the central result of this Letter. The quadrivector $\mathbf{J}^\sigma$, accounting for all quadratic combinations of the helicity-dependent responses $\{\gamma^\sigma, \beta^\sigma\}$, can be captured from a measurement of the Stokes vector $\mathbf{S}^\sigma (kr, \theta)$  evaluated at a single scattering angle $\theta$ in the far-field. As Eq.~\eqref{compact} shows, one only needs to apply a simple 4x4 matrix ${\til{U}^\sigma (kr, \theta)}$  to the measurement of $\mathbf{S}^\sigma (kr, \theta)$ to obtain $\mathbf{J}^\sigma$. It is important to note that $\mathbf{J}^\sigma$ does not depend on the amplitude of the incident field $E_0$, which cancels out.
In addition to this, the optical distance from the center of the sample to the observational point, denoted by $kr$, does not modify $\mathbf{J}^\sigma$ as long as the Stokes vector is measured in the far-field. This is because the Stokes vector is defined in the far-field and thus $\mathbf{S}^\sigma (kr, \theta) \propto (kr)^{-2}$, which cancels the $(kr)^2$ in Eq.~\eqref{compact}. Moreover, we anticipate that knowing $N$ is not needed as it will cancel out in the definition of our novel chirality measure. 

We now discuss how to use Eq.~\eqref{compact} in the laboratory.
To start with, we note that the scattering angle can be freely chosen in the interval $0< \theta < \pi $, where $\csc^4 \theta\neq 0$. This holds since $\mathbf{J}^\sigma$ does not depend on the choice of $\theta$. Nevertheless, one must  avoid any propagation direction where the incident field has a component. Otherwise, the total field measured in the optical laboratory will be the sum of the scattered and incident fields, thus invalidating Eq.~\eqref{compact}. Moreover,  it is important to note that the angle $\varphi$ does not play any role in Eq.~\eqref{compact} since the system is symmetrical in azimuth. Thus, $\varphi$ can be chosen freely in the optical laboratory.  

\begin{figure*}
    \centering
\includegraphics[width=0.975\linewidth]{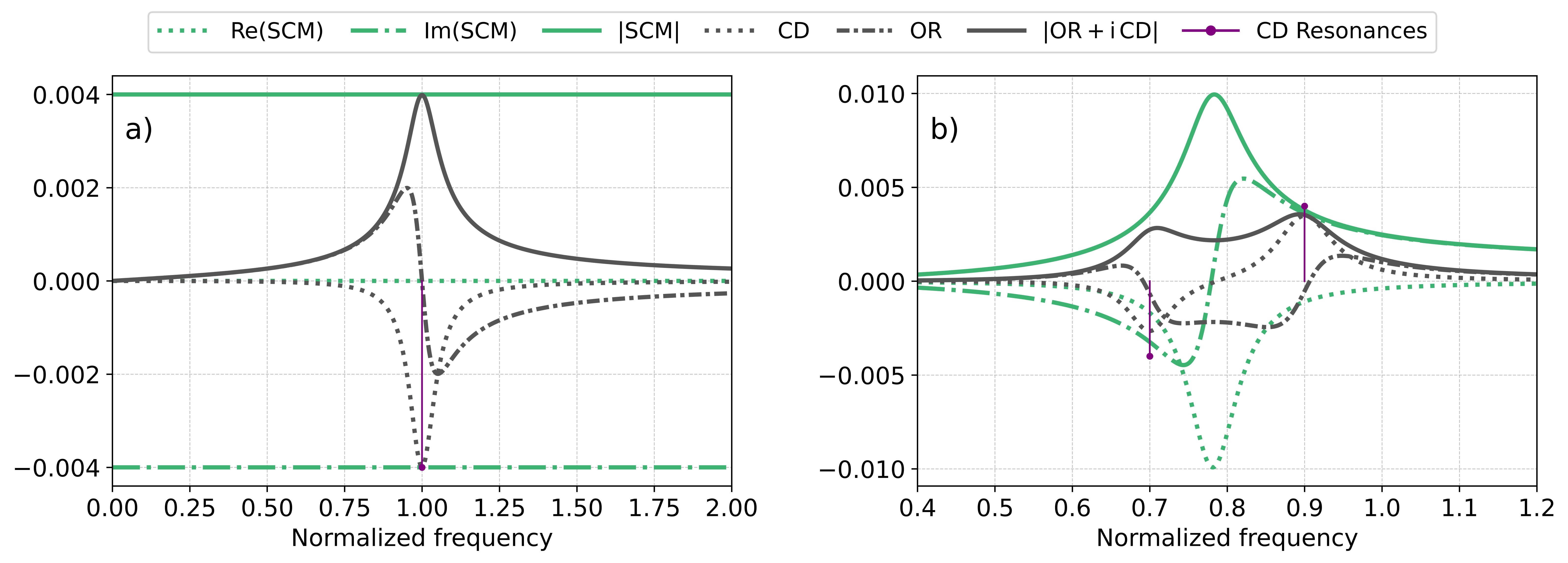}
    \caption{Stokes chirality measure (SCM), CD, and OR, calculated with \Eq{SCM_new}, \Eq{eq:CD}, and \Eq{eq:OR}, respectively. The CD of each isolated resonance computed with \Eq{CD} is shown as purple stem bars.  The frequency axis is normalized to 500THz. All the quantities are dimensionless. a) Single CD resonance at $\omega/\omega_{\rm{res}} = 1$ with $s_{\rm{res}} = 1$. b) Two CD resonances at $\omega/\omega_{\rm{res}} = 0.7$ with $s_{\rm{res}} = 1$ and $\omega/\omega_{\rm{res}} = 0.9$ with $s_{\rm{res}} = -1$. In all cases, we consider $\tau_{\rm{res}}=$1e-6, $A_{\text{res}}=2.2727$e-34, and $\gamma_{\rm{res}}/w_{\rm{res}} = 0.1$. }
    \label{fig:M}
\end{figure*}

Additionally and regarding the fidelity of our method, we highlight that it is robust upon two measurements of the Stokes vector at distinct observational points. That is, if we apply Eq.~\eqref{compact} for two different angles $\theta_1$ and $\theta_2$ and obtain the same $\mathbf{J}^\sigma$, then our method is reliable. Otherwise, it indicates that the setting we assumed is not valid. Such a practical consistency check is  valuable in the optical laboratory.

Next, we show that Eq.~\eqref{compact} can be used to formulate a novel method for detecting the chirality of spherical objects.

\emph{A novel chiroptical spectroscopy technique.}---
We start this section by considering the difference, $\mathbf{D} = \mathbf{J}^+ - \mathbf{J}^-$, and the sum, $\mathbf{S} = (\mathbf{J}^+ + \mathbf{J}^-)/2$, quadrivectors: 
\begin{equation}\label{D-}
\mathbf{D} = {\mathbf{J}^+ - \mathbf{J}^-} =N 
{\begin{pmatrix}
 4\Im \{{\alpha}^*_{\rm{ee}} {\alpha}_{\rm{em}} \} \\
 4\Im \{{\alpha}^*_{\rm{mm}} {\alpha}_{\rm{em}} \}\\ 2\Im \{ ({\alpha}_{\rm{ee}} + {\alpha}_{\rm{mm}})^* {\alpha}_{\rm{em}} \}\\
   2\Re \{ ({\alpha}_{\rm{ee}} - {\alpha}_{\rm{mm}})^* {\alpha}_{\rm{em}} \}
    \end{pmatrix}},
\end{equation}
\begin{equation}\label{s}
\mathbf{S} = \frac{\mathbf{J}^+ + \mathbf{J}^-}{2} =N 
{\begin{pmatrix}
 |{\alpha}_{\rm{ee}}|^2 +  |{\alpha}_{\rm{em}}|^2 \\
  |{\alpha}_{\rm{mm}}|^2 +  |{\alpha}_{\rm{em}}|^2\\ \Re \{{\alpha}_{\rm{ee}} {\alpha}_{\rm{mm}}^* \} + |{\alpha}_{\rm{em}}|^2 \\
  \Im \{{\alpha}_{\rm{ee}} {\alpha}_{\rm{mm}}^* \}
    \end{pmatrix}}.
\end{equation}
We note that if any of the four components of $\mathbf{D}$ is different from zero, the object is chiral, namely, $|\alpha_{\rm{em}}| \neq 0$. There is, however, one exception to this rule: the first Kerker condition $\alpha_K = \alpha_{\rm{ee}} = \alpha_{\rm{mm}}$~\cite{kerker1983electromagnetic, olmos2020unveiling, olmos2022helicity} combined with  $\Im \{\alpha_K \alpha^*_{\rm{em}} \}  = 0$. 
In this unconventional case, additional observables may be required to determine the presence of chirality at a fixed frequency $\omega$ of the incident electromagnetic field. Moreover, if $\alpha_{\rm{em}} = 0$, then we recover the Stokes vector method for achiral objects, firstly introduced in Refs.~\cite{olmos2024revealing, olmos2024solving}


Having noted these points, we now discuss the underlying physics for $\D \neq 0$. If we know that the sample is a solution containing the two versions of a given chiral object, we can further detect which enantiomer is more abundant in the solution. This is evidenced by the change in sign of $\mathbf{D}$ depending on the specific enantiomer. Note that for two enantiomers $i$ and $j$ of the same chiral object, we have $\alpha^i_{\rm{em}} = -\alpha^j_{\rm{em}}$.

Many different chirality measures can be derived from $\mathbf{D}$. We now present one of the most interesting ones. We call our quantity the Stokes Chirality Measure (SCM), defined as:
\begin{equation} \label{SCM}
\text{SCM}= 2 \left[\frac{\mathbf{D}(4)-\ii \mathbf{D}(1)/2}{\mathbf{S}(1)+\mathbf{S}(2)} \right].
\end{equation}
{We will assume noiseless measurements for the forthcoming calculations.} Expanding Eq.~\eqref{SCM} using Eqs.~\eqref{gamma}-\eqref{s}, we get
\begin{equation}
\label{SCM_new}
\text{SCM}=4 \left[\frac{\Re\{(\alphaee-\alphamm)^*\alphaem\}-\ii\Im\{\alphaee^*\alphaem\}}{|\alphaee|^2+2|\alphaem|^2+|\alphamm|^2} \right].
\end{equation}
The SCM is unitless due to the division by $\mathbf{S}(1)+\mathbf{S}(2)$, which can be seen as the total ``size squared'' of the polarizability tensor. To be more explicit, if $\mathbf{S}(1)+\mathbf{S}(2) = 0$ then the object is transparent to any incident field~\cite{olmos2024spheres}. Here $\mathbf{S}(1)$ denotes the first component of $\mathbf{S}$, namely, $\mathbf{S}(1) = N \left(|\alpha_{\rm{ee}}|^2 + |\alpha_{\rm{em}}|^2  \right)$. In this vein, we note that the imaginary part of the SCM can also be obtained as $2(\mathbf{D}(3)-\mathbf{D}(2)/2)/(\mathbf{S}(1)+\mathbf{S}(2))$. An average of the two expressions should reduce the noise when the SCM is obtained from laboratory measurements.

We now compare Eq.~\eqref{SCM_new} with the standard OR and CD spectroscopy measurements.

\emph{Results.}---
A model of the dipolar chiro-optical response is needed for computing examples of the chirality measures. We will use the quasi-static model for the dipolar resonance of a chiral object from~\cite[Eq.~(20)]{Sersic2011}, which assumes that the electric and magnetic dipolar moments originate from the same current distribution. Under such assumptions, the frequency-dependent polarizability of each resonance can be written as:
\begin{equation}
	\label{eq:nonsi}
\begin{split}
& \begin{pmatrix}
{\alpha}_{\rm{ee}}(\omega) & {\alpha}_{\rm{em}}(\omega)  \\
-{\alpha}_{\rm{em}}(\omega) & {\alpha}_{\rm{mm}}(\omega)
\end{pmatrix}= A(\omega)
	\begin{pmatrix}1&\ii s_{\text{res}}\sqrt{\tau_{\text{res}}}\\-\ii s_{\text{res}}\sqrt{\tau_{\text{res}}}&\tau_{\text{res}}\end{pmatrix},
 \end{split}
\end{equation}
where 
\begin{equation} \nonumber
A(\omega) = \frac{A_\text{res}}{1-\left(\frac{\omega}{\omega_{\text{res}}}\right)^2-\ii\frac{\gamma_{\text{res}}\omega}{\omega^2_{\text{res}}}}.    
\end{equation}
Here $\omega_{\text{res}}$ is the resonance frequency, $\gamma_{\text{res}}$ determines the linewidth, $\tau_{\text{res}}\ge 0$, and $s_{\text{res}}=\{-1,+1\}$ determines the handedness of the resonance. 
Opposite enantiomers will feature opposite signs of $s_{\text{res}}$. When $\tau_{\text{res}}=0$, the resonance is purely electric and there is no chiral response. Now, the CD of an isolated resonance at $\omega=\omega_{\text{res}}$ is \cite[Eq.~(12)]{Scott2020}:
\begin{equation}
\label{CD}
\text{CD}(\omega_{\text{res}})=-s_{\text{res}}\frac{4\sqrt{\tau_{\text{res}}}}{1+\tau_{\text{res}}},
\end{equation}
where the sign has been flipped for a more convenient comparison with the typical expression of CD in a transmission measurement.  
Eq.~(\ref{CD}) helps in selecting relevant values of $\tau_{\text{res}}$. For example, for chiral molecules the largest values of CD in electronic, vibrational, and rotational CD spectroscopy are of the order 10$^{-1}$, 10$^{-3}$, and 10$^{-5}$, respectively, and the typical values are at least an order of magnitude lower in each case \cite{Mason1963,Jasco2011,Salzman1991}. 
We will consider systems with multiple resonances, and add the polarizabilities of each resonance 
to obtain the total polarizability tensor of the object.

The following expressions are often used for the CD and OR in the forward direction (see e.g. \cite[Eqs.~(1.2.5,1.2.8)]{Barron2004}):
\begin{align}
\label{eq:CD}
	\text{CD}(\omega)&=-2\frac{e^{-k(\omega)L\Im\{n_+(\omega)\}}-e^{-k(\omega)L\Im\{n_-(\omega)\}}}{e^{-k(\omega)L\Im\{n_+(\omega)\}}+e^{-k(\omega)L\Im\{n_-(\omega)\}}},\\
\label{eq:OR}
	\text{OR}(\omega)&=k(\omega)L\Re\{n_+(\omega)-n_-(\omega)\},
\end{align}
where $L$ is the optical path of the illuminating beam inside the sample, $k(\omega)=\omega/c$, and $n_\pm(\omega)$ are the effective refractive indexes for the two helicities of the incident field. Such refractive indexes can be computed using homogenization techniques as (see e.g.~\cite[Eq.~(6.127)]{Lindell1994}):
\begin{equation} \nonumber
\begin{split}&n_\pm(\omega)=\sqrt{\epsilon_r(\omega)\mu_r(\omega)}\pm \kappa(\omega),\
\epsilon_r(\omega)=1+\frac{4\pi}{3}\frac{N}{V_\text{int}}\alpha_{\rm{ee}}(\omega),\\ \nonumber
&\mu_r(\omega)=1+\frac{4\pi}{3}\frac{N}{V_\text{int}}\alpha_{\rm{mm}}(\omega),\
\kappa(\omega)=\frac{-\ii 4\pi}{3}\frac{N}{V_\text{int}}\alpha_{\rm{em}}(\omega),\\
\end{split}
\end{equation}
where  $V_\text{int}$ is the volume that they occupy. We have assumed for simplicity, but without loss of generality, that the objects are in vacuum.
We also assume $L=$\SI{0.01}{\meter}, a beam spot size of area 1mm$^2$, with which $V_{\text{int}}=10^{-8}$m$^3$, and a concentration of objects of 1mg/ml. The value of $N$ depends then on the molecular mass, and we use $N=2$e19.

The $A_{\text{res}}$ in \Eq{eq:nonsi} remain to be fixed. For a system with a single resonance, $A_{\text{res}}$ does not affect the SCM because it cancels out. Its value is, however, needed for computing CD and OR. We have chosen $A_{\text{res}}=2.2727$e-34 with the following reasoning. For a system with a single resonance, such value makes the result of Eq.~\eqref{CD} equal to the result of Eq.~\eqref{eq:CD} at $\omega_{\text{res}}=(2\pi)500e12$, for the relevant values of $\tau_{\text{res}}$. Incidentally, the imaginary part of SCM($\omega_{\text{res}}$) coincides as well, as can be seen in Fig.~\ref{fig:M}(a). We then keep $A_{\text{res}}=2.2727$e-34 for all the resonances in the  calculations presented in Fig.~\ref{fig:M}. 

Figure~\ref{fig:M} shows that the SCM provides a signature of chirality that is different from the typical CD and OR signatures. The case of a single resonance in Fig.~\ref{fig:M}(a) is the extreme illustration of this point. The frequency-independent behavior of the SCM is, however, a consequence of the model that we use for the polarizability of a chiral resonance, and we do not expect this to be a general property for the SCM of isolated resonances. Figure~\ref{fig:M}(b) shows the results for the case of two resonances. In this case, while still different, the real and imaginary parts of the SCM are closer to the OR and the CD, respectively. Unsigned measures of chirality, obtained as the absolute values of SCM, and OR+$\ii$CD, respectively, are shown in Fig.~\ref{fig:M} as well. The unsigned measures enable the detection of chirality. We see from the definitions of their real and imaginary parts, in \Eq{SCM_new}, \Eq{eq:CD}, and \Eq{eq:OR}, that their sign flips with the sign of $\alphaem(\omega)$. Therefore, such real and imaginary parts are suitable for identifying enantiomers. 

As we previously anticipated, there is an important difference between SCM, and CD and OR: The value of SCM is independent of the number of excited objects; it only depends on the polarizabilities [\Eq{SCM_new}]. This is an advantage over the CD and OR, which depend on the concentration of the species in the solution and the path length [\Eq{eq:CD}, \Eq{eq:OR}].


\emph{Conclusion.}--- We introduce a spectroscopic technique for detecting the presence and sign of the chirality of spherical objects based on measuring the Stokes parameters at non-forward directions. Crucially, such measurements are experimentally feasible for chiral molecules~\cite{Vidal2015} and achiral objects~\cite{negoro2023helicity, sugimoto2020mie}. Our approach opens a new route to probe chirality beyond the constraints of conventional techniques such as CD and OR, potentially redefining the spectroscopic toolbox for chiral light–matter interactions at the nanoscale.

\bibliography{Bib_tesis}

\begin{thebibliography}{44}%
\makeatletter
\providecommand \@ifxundefined [1]{%
 \@ifx{#1\undefined}
}%
\providecommand \@ifnum [1]{%
 \ifnum #1\expandafter \@firstoftwo
 \else \expandafter \@secondoftwo
 \fi
}%
\providecommand \@ifx [1]{%
 \ifx #1\expandafter \@firstoftwo
 \else \expandafter \@secondoftwo
 \fi
}%
\providecommand \natexlab [1]{#1}%
\providecommand \enquote  [1]{``#1''}%
\providecommand \bibnamefont  [1]{#1}%
\providecommand \bibfnamefont [1]{#1}%
\providecommand \citenamefont [1]{#1}%
\providecommand \href@noop [0]{\@secondoftwo}%
\providecommand \href [0]{\begingroup \@sanitize@url \@href}%
\providecommand \@href[1]{\@@startlink{#1}\@@href}%
\providecommand \@@href[1]{\endgroup#1\@@endlink}%
\providecommand \@sanitize@url [0]{\catcode `\\12\catcode `\$12\catcode
  `\&12\catcode `\#12\catcode `\^12\catcode `\_12\catcode `\%12\relax}%
\providecommand \@@startlink[1]{}%
\providecommand \@@endlink[0]{}%
\providecommand \url  [0]{\begingroup\@sanitize@url \@url }%
\providecommand \@url [1]{\endgroup\@href {#1}{\urlprefix }}%
\providecommand \urlprefix  [0]{URL }%
\providecommand \Eprint [0]{\href }%
\providecommand \doibase [0]{http://dx.doi.org/}%
\providecommand \selectlanguage [0]{\@gobble}%
\providecommand \bibinfo  [0]{\@secondoftwo}%
\providecommand \bibfield  [0]{\@secondoftwo}%
\providecommand \translation [1]{[#1]}%
\providecommand \BibitemOpen [0]{}%
\providecommand \bibitemStop [0]{}%
\providecommand \bibitemNoStop [0]{.\EOS\space}%
\providecommand \EOS [0]{\spacefactor3000\relax}%
\providecommand \BibitemShut  [1]{\csname bibitem#1\endcsname}%
\let\auto@bib@innerbib\@empty
\bibitem [{\citenamefont {Lenz}\ and\ \citenamefont
  {Knapp}(1962)}]{lenz1962thalidomide}%
  \BibitemOpen
  \bibfield  {author} {\bibinfo {author} {\bibfnamefont {W.}~\bibnamefont
  {Lenz}}\ and\ \bibinfo {author} {\bibfnamefont {K.}~\bibnamefont {Knapp}},\
  }\href@noop {} {\bibfield  {journal} {\bibinfo  {journal} {Archives of
  Environmental Health: An International Journal}\ }\textbf {\bibinfo {volume}
  {5}},\ \bibinfo {pages} {14} (\bibinfo {year} {1962})}\BibitemShut {NoStop}%
\bibitem [{\citenamefont {Fernandez-Corbaton}\ \emph
  {et~al.}(2013)\citenamefont {Fernandez-Corbaton}, \citenamefont
  {Zambrana-Puyalto}, \citenamefont {Tischler}, \citenamefont {Vidal},
  \citenamefont {Juan},\ and\ \citenamefont
  {Molina-Terriza}}]{fernandez2013electromagnetic}%
  \BibitemOpen
  \bibfield  {author} {\bibinfo {author} {\bibfnamefont {I.}~\bibnamefont
  {Fernandez-Corbaton}}, \bibinfo {author} {\bibfnamefont {X.}~\bibnamefont
  {Zambrana-Puyalto}}, \bibinfo {author} {\bibfnamefont {N.}~\bibnamefont
  {Tischler}}, \bibinfo {author} {\bibfnamefont {X.}~\bibnamefont {Vidal}},
  \bibinfo {author} {\bibfnamefont {M.~L.}\ \bibnamefont {Juan}}, \ and\
  \bibinfo {author} {\bibfnamefont {G.}~\bibnamefont {Molina-Terriza}},\
  }\href@noop {} {\bibfield  {journal} {\bibinfo  {journal} {Phys. Rev. Lett.}\
  }\textbf {\bibinfo {volume} {111}},\ \bibinfo {pages} {060401} (\bibinfo
  {year} {2013})}\BibitemShut {NoStop}%
\bibitem [{\citenamefont {Barron}(2009)}]{barron2009molecular}%
  \BibitemOpen
  \bibfield  {author} {\bibinfo {author} {\bibfnamefont {L.~D.}\ \bibnamefont
  {Barron}},\ }\href@noop {} {\emph {\bibinfo {title} {Molecular light
  scattering and optical activity}}}\ (\bibinfo  {publisher} {Cambridge
  University Press},\ \bibinfo {year} {2009})\BibitemShut {NoStop}%
\bibitem [{\citenamefont {Adhikari}\ and\ \citenamefont
  {Orrit}(2022)}]{adhikari2022optically}%
  \BibitemOpen
  \bibfield  {author} {\bibinfo {author} {\bibfnamefont {S.}~\bibnamefont
  {Adhikari}}\ and\ \bibinfo {author} {\bibfnamefont {M.}~\bibnamefont
  {Orrit}},\ }\href@noop {} {\bibfield  {journal} {\bibinfo  {journal} {Acs
  Photonics}\ }\textbf {\bibinfo {volume} {9}},\ \bibinfo {pages} {3486}
  (\bibinfo {year} {2022})}\BibitemShut {NoStop}%
\bibitem [{\citenamefont {Tang}\ and\ \citenamefont
  {Cohen}(2010)}]{tang2010optical}%
  \BibitemOpen
  \bibfield  {author} {\bibinfo {author} {\bibfnamefont {Y.}~\bibnamefont
  {Tang}}\ and\ \bibinfo {author} {\bibfnamefont {A.~E.}\ \bibnamefont
  {Cohen}},\ }\href@noop {} {\bibfield  {journal} {\bibinfo  {journal}
  {Physical review letters}\ }\textbf {\bibinfo {volume} {104}},\ \bibinfo
  {pages} {163901} (\bibinfo {year} {2010})}\BibitemShut {NoStop}%
\bibitem [{\citenamefont {Sersic}\ \emph {et~al.}(2011)\citenamefont {Sersic},
  \citenamefont {Tuambilangana}, \citenamefont {Kampfrath},\ and\ \citenamefont
  {Koenderink}}]{Sersic2011}%
  \BibitemOpen
  \bibfield  {author} {\bibinfo {author} {\bibfnamefont {I.}~\bibnamefont
  {Sersic}}, \bibinfo {author} {\bibfnamefont {C.}~\bibnamefont
  {Tuambilangana}}, \bibinfo {author} {\bibfnamefont {T.}~\bibnamefont
  {Kampfrath}}, \ and\ \bibinfo {author} {\bibfnamefont {A.~F.}\ \bibnamefont
  {Koenderink}},\ }\href {\doibase 10.1103/PhysRevB.83.245102} {\bibfield
  {journal} {\bibinfo  {journal} {Phys. Rev. B}\ }\textbf {\bibinfo {volume}
  {83}},\ \bibinfo {pages} {245102} (\bibinfo {year} {2011})}\BibitemShut
  {NoStop}%
\bibitem [{\citenamefont {Scott}\ \emph {et~al.}(2020)\citenamefont {Scott},
  \citenamefont {Garcia-Santiago}, \citenamefont {Beutel}, \citenamefont
  {Rockstuhl}, \citenamefont {Wegener},\ and\ \citenamefont
  {Fernandez-Corbaton}}]{Scott2020}%
  \BibitemOpen
  \bibfield  {author} {\bibinfo {author} {\bibfnamefont {P.}~\bibnamefont
  {Scott}}, \bibinfo {author} {\bibfnamefont {X.}~\bibnamefont
  {Garcia-Santiago}}, \bibinfo {author} {\bibfnamefont {D.}~\bibnamefont
  {Beutel}}, \bibinfo {author} {\bibfnamefont {C.}~\bibnamefont {Rockstuhl}},
  \bibinfo {author} {\bibfnamefont {M.}~\bibnamefont {Wegener}}, \ and\
  \bibinfo {author} {\bibfnamefont {I.}~\bibnamefont {Fernandez-Corbaton}},\
  }\href {\doibase 10.1063/5.0025006} {\bibfield  {journal} {\bibinfo
  {journal} {Applied Physics Reviews}\ }\textbf {\bibinfo {volume} {7}},\
  \bibinfo {pages} {041413} (\bibinfo {year} {2020})}\BibitemShut {NoStop}%
\bibitem [{\citenamefont {Bohren}(1974)}]{bohren1974light}%
  \BibitemOpen
  \bibfield  {author} {\bibinfo {author} {\bibfnamefont {C.~F.}\ \bibnamefont
  {Bohren}},\ }\href@noop {} {\bibfield  {journal} {\bibinfo  {journal}
  {Chemical Physics Letters}\ }\textbf {\bibinfo {volume} {29}},\ \bibinfo
  {pages} {458} (\bibinfo {year} {1974})}\BibitemShut {NoStop}%
\bibitem [{\citenamefont {Yokota}\ \emph {et~al.}(2001)\citenamefont {Yokota},
  \citenamefont {He},\ and\ \citenamefont {Takenaka}}]{yokota2001scattering}%
  \BibitemOpen
  \bibfield  {author} {\bibinfo {author} {\bibfnamefont {M.}~\bibnamefont
  {Yokota}}, \bibinfo {author} {\bibfnamefont {S.}~\bibnamefont {He}}, \ and\
  \bibinfo {author} {\bibfnamefont {T.}~\bibnamefont {Takenaka}},\ }\href@noop
  {} {\bibfield  {journal} {\bibinfo  {journal} {Journal of the Optical Society
  of America A}\ }\textbf {\bibinfo {volume} {18}},\ \bibinfo {pages} {1681}
  (\bibinfo {year} {2001})}\BibitemShut {NoStop}%
\bibitem [{\citenamefont {Canaguier-Durand}\ \emph {et~al.}(2013)\citenamefont
  {Canaguier-Durand}, \citenamefont {Hutchison}, \citenamefont {Genet},\ and\
  \citenamefont {Ebbesen}}]{canaguier2013mechanical}%
  \BibitemOpen
  \bibfield  {author} {\bibinfo {author} {\bibfnamefont {A.}~\bibnamefont
  {Canaguier-Durand}}, \bibinfo {author} {\bibfnamefont {J.~A.}\ \bibnamefont
  {Hutchison}}, \bibinfo {author} {\bibfnamefont {C.}~\bibnamefont {Genet}}, \
  and\ \bibinfo {author} {\bibfnamefont {T.~W.}\ \bibnamefont {Ebbesen}},\
  }\href@noop {} {\bibfield  {journal} {\bibinfo  {journal} {New Journal of
  Physics}\ }\textbf {\bibinfo {volume} {15}},\ \bibinfo {pages} {123037}
  (\bibinfo {year} {2013})}\BibitemShut {NoStop}%
\bibitem [{\citenamefont {Hayat}\ \emph {et~al.}(2015)\citenamefont {Hayat},
  \citenamefont {Mueller},\ and\ \citenamefont {Capasso}}]{hayat2015lateral}%
  \BibitemOpen
  \bibfield  {author} {\bibinfo {author} {\bibfnamefont {A.}~\bibnamefont
  {Hayat}}, \bibinfo {author} {\bibfnamefont {J.~B.}\ \bibnamefont {Mueller}},
  \ and\ \bibinfo {author} {\bibfnamefont {F.}~\bibnamefont {Capasso}},\
  }\href@noop {} {\bibfield  {journal} {\bibinfo  {journal} {Proceedings of the
  National Academy of Sciences}\ }\textbf {\bibinfo {volume} {112}},\ \bibinfo
  {pages} {13190} (\bibinfo {year} {2015})}\BibitemShut {NoStop}%
\bibitem [{\citenamefont {Li}\ \emph {et~al.}(2019)\citenamefont {Li},
  \citenamefont {Yan}, \citenamefont {Zhang}, \citenamefont {Liang},
  \citenamefont {Zhang},\ and\ \citenamefont {Yao}}]{li2019optical}%
  \BibitemOpen
  \bibfield  {author} {\bibinfo {author} {\bibfnamefont {M.}~\bibnamefont
  {Li}}, \bibinfo {author} {\bibfnamefont {S.}~\bibnamefont {Yan}}, \bibinfo
  {author} {\bibfnamefont {Y.}~\bibnamefont {Zhang}}, \bibinfo {author}
  {\bibfnamefont {Y.}~\bibnamefont {Liang}}, \bibinfo {author} {\bibfnamefont
  {P.}~\bibnamefont {Zhang}}, \ and\ \bibinfo {author} {\bibfnamefont
  {B.}~\bibnamefont {Yao}},\ }\href@noop {} {\bibfield  {journal} {\bibinfo
  {journal} {Physical Review A}\ }\textbf {\bibinfo {volume} {99}},\ \bibinfo
  {pages} {033825} (\bibinfo {year} {2019})}\BibitemShut {NoStop}%
\bibitem [{\citenamefont {Klimov}\ \emph {et~al.}(2014)\citenamefont {Klimov},
  \citenamefont {Zabkov}, \citenamefont {Pavlov},\ and\ \citenamefont
  {Guzatov}}]{klimov2014eigen}%
  \BibitemOpen
  \bibfield  {author} {\bibinfo {author} {\bibfnamefont {V.}~\bibnamefont
  {Klimov}}, \bibinfo {author} {\bibfnamefont {I.}~\bibnamefont {Zabkov}},
  \bibinfo {author} {\bibfnamefont {A.}~\bibnamefont {Pavlov}}, \ and\ \bibinfo
  {author} {\bibfnamefont {D.}~\bibnamefont {Guzatov}},\ }\href@noop {}
  {\bibfield  {journal} {\bibinfo  {journal} {Optics Express}\ }\textbf
  {\bibinfo {volume} {22}},\ \bibinfo {pages} {18564} (\bibinfo {year}
  {2014})}\BibitemShut {NoStop}%
\bibitem [{\citenamefont {Wang}\ and\ \citenamefont
  {Chan}(2014)}]{wang2014lateral}%
  \BibitemOpen
  \bibfield  {author} {\bibinfo {author} {\bibfnamefont {S.}~\bibnamefont
  {Wang}}\ and\ \bibinfo {author} {\bibfnamefont {C.~T.}\ \bibnamefont
  {Chan}},\ }\href@noop {} {\bibfield  {journal} {\bibinfo  {journal} {Nature
  communications}\ }\textbf {\bibinfo {volume} {5}},\ \bibinfo {pages} {3307}
  (\bibinfo {year} {2014})}\BibitemShut {NoStop}%
\bibitem [{\citenamefont {Kamandi}\ \emph {et~al.}(2017)\citenamefont
  {Kamandi}, \citenamefont {Albooyeh}, \citenamefont {Guclu}, \citenamefont
  {Veysi}, \citenamefont {Zeng}, \citenamefont {Wickramasinghe},\ and\
  \citenamefont {Capolino}}]{kamandi2017enantiospecific}%
  \BibitemOpen
  \bibfield  {author} {\bibinfo {author} {\bibfnamefont {M.}~\bibnamefont
  {Kamandi}}, \bibinfo {author} {\bibfnamefont {M.}~\bibnamefont {Albooyeh}},
  \bibinfo {author} {\bibfnamefont {C.}~\bibnamefont {Guclu}}, \bibinfo
  {author} {\bibfnamefont {M.}~\bibnamefont {Veysi}}, \bibinfo {author}
  {\bibfnamefont {J.}~\bibnamefont {Zeng}}, \bibinfo {author} {\bibfnamefont
  {K.}~\bibnamefont {Wickramasinghe}}, \ and\ \bibinfo {author} {\bibfnamefont
  {F.}~\bibnamefont {Capolino}},\ }\href@noop {} {\bibfield  {journal}
  {\bibinfo  {journal} {Physical Review Applied}\ }\textbf {\bibinfo {volume}
  {8}},\ \bibinfo {pages} {064010} (\bibinfo {year} {2017})}\BibitemShut
  {NoStop}%
\bibitem [{\citenamefont {Sifat}\ \emph {et~al.}(2022)\citenamefont {Sifat},
  \citenamefont {Capolino},\ and\ \citenamefont {Potma}}]{sifat2022force}%
  \BibitemOpen
  \bibfield  {author} {\bibinfo {author} {\bibfnamefont {A.~A.}\ \bibnamefont
  {Sifat}}, \bibinfo {author} {\bibfnamefont {F.}~\bibnamefont {Capolino}}, \
  and\ \bibinfo {author} {\bibfnamefont {E.~O.}\ \bibnamefont {Potma}},\
  }\href@noop {} {\bibfield  {journal} {\bibinfo  {journal} {ACS Photonics}\
  }\textbf {\bibinfo {volume} {9}},\ \bibinfo {pages} {2660} (\bibinfo {year}
  {2022})}\BibitemShut {NoStop}%
\bibitem [{\citenamefont {Ali}\ \emph {et~al.}(2023)\citenamefont {Ali},
  \citenamefont {Pinheiro}, \citenamefont {Dutra}, \citenamefont {Alegre},\
  and\ \citenamefont {Wiederhecker}}]{ali2023enantioselective}%
  \BibitemOpen
  \bibfield  {author} {\bibinfo {author} {\bibfnamefont {R.}~\bibnamefont
  {Ali}}, \bibinfo {author} {\bibfnamefont {F.}~\bibnamefont {Pinheiro}},
  \bibinfo {author} {\bibfnamefont {R.}~\bibnamefont {Dutra}}, \bibinfo
  {author} {\bibfnamefont {T.~M.}\ \bibnamefont {Alegre}}, \ and\ \bibinfo
  {author} {\bibfnamefont {G.}~\bibnamefont {Wiederhecker}},\ }\href@noop {}
  {\bibfield  {journal} {\bibinfo  {journal} {Physical Review A}\ }\textbf
  {\bibinfo {volume} {108}},\ \bibinfo {pages} {043704} (\bibinfo {year}
  {2023})}\BibitemShut {NoStop}%
\bibitem [{\citenamefont {Kim}\ and\ \citenamefont
  {Park}(2022)}]{kim2022molecular}%
  \BibitemOpen
  \bibfield  {author} {\bibinfo {author} {\bibfnamefont {T.}~\bibnamefont
  {Kim}}\ and\ \bibinfo {author} {\bibfnamefont {Q.-H.}\ \bibnamefont {Park}},\
  }\href@noop {} {\bibfield  {journal} {\bibinfo  {journal} {Nanophotonics}\
  }\textbf {\bibinfo {volume} {11}},\ \bibinfo {pages} {1897} (\bibinfo {year}
  {2022})}\BibitemShut {NoStop}%
\bibitem [{\citenamefont {Sadrara}\ and\ \citenamefont
  {Miri}(2023)}]{sadrara2023large}%
  \BibitemOpen
  \bibfield  {author} {\bibinfo {author} {\bibfnamefont {M.}~\bibnamefont
  {Sadrara}}\ and\ \bibinfo {author} {\bibfnamefont {M.}~\bibnamefont {Miri}},\
  }\href@noop {} {\bibfield  {journal} {\bibinfo  {journal} {The Journal of
  Physical Chemistry C}\ }\textbf {\bibinfo {volume} {128}},\ \bibinfo {pages}
  {218} (\bibinfo {year} {2023})}\BibitemShut {NoStop}%
\bibitem [{\citenamefont {Shi}\ \emph {et~al.}(2020)\citenamefont {Shi},
  \citenamefont {Zhu}, \citenamefont {Zhang}, \citenamefont {Mazzulla},
  \citenamefont {Tsai}, \citenamefont {Ding}, \citenamefont {Liu},
  \citenamefont {Cipparrone}, \citenamefont {S{\'a}enz},\ and\ \citenamefont
  {Qiu}}]{shi2020chirality}%
  \BibitemOpen
  \bibfield  {author} {\bibinfo {author} {\bibfnamefont {Y.}~\bibnamefont
  {Shi}}, \bibinfo {author} {\bibfnamefont {T.}~\bibnamefont {Zhu}}, \bibinfo
  {author} {\bibfnamefont {T.}~\bibnamefont {Zhang}}, \bibinfo {author}
  {\bibfnamefont {A.}~\bibnamefont {Mazzulla}}, \bibinfo {author}
  {\bibfnamefont {D.~P.}\ \bibnamefont {Tsai}}, \bibinfo {author}
  {\bibfnamefont {W.}~\bibnamefont {Ding}}, \bibinfo {author} {\bibfnamefont
  {A.~Q.}\ \bibnamefont {Liu}}, \bibinfo {author} {\bibfnamefont
  {G.}~\bibnamefont {Cipparrone}}, \bibinfo {author} {\bibfnamefont {J.~J.}\
  \bibnamefont {S{\'a}enz}}, \ and\ \bibinfo {author} {\bibfnamefont {C.-W.}\
  \bibnamefont {Qiu}},\ }\href@noop {} {\bibfield  {journal} {\bibinfo
  {journal} {Light: Science \& Applications}\ }\textbf {\bibinfo {volume}
  {9}},\ \bibinfo {pages} {1} (\bibinfo {year} {2020})}\BibitemShut {NoStop}%
\bibitem [{\citenamefont {Mohammadi}\ \emph {et~al.}(2023)\citenamefont
  {Mohammadi}, \citenamefont {Raziman},\ and\ \citenamefont
  {Curto}}]{mohammadi2023nanophotonic}%
  \BibitemOpen
  \bibfield  {author} {\bibinfo {author} {\bibfnamefont {E.}~\bibnamefont
  {Mohammadi}}, \bibinfo {author} {\bibfnamefont {T.}~\bibnamefont {Raziman}},
  \ and\ \bibinfo {author} {\bibfnamefont {A.~G.}\ \bibnamefont {Curto}},\
  }\href@noop {} {\bibfield  {journal} {\bibinfo  {journal} {Nano Letters}\
  }\textbf {\bibinfo {volume} {23}},\ \bibinfo {pages} {3978} (\bibinfo {year}
  {2023})}\BibitemShut {NoStop}%
\bibitem [{\citenamefont {Mart{\'\i}nez-Romeu}\ \emph
  {et~al.}(2024)\citenamefont {Mart{\'\i}nez-Romeu}, \citenamefont {Diez},
  \citenamefont {Golat}, \citenamefont {Rodr{\'\i}guez-Fortu{\~n}o},\ and\
  \citenamefont {Mart{\'\i}nez}}]{martinez2024longitudinal}%
  \BibitemOpen
  \bibfield  {author} {\bibinfo {author} {\bibfnamefont {J.}~\bibnamefont
  {Mart{\'\i}nez-Romeu}}, \bibinfo {author} {\bibfnamefont {I.}~\bibnamefont
  {Diez}}, \bibinfo {author} {\bibfnamefont {S.}~\bibnamefont {Golat}},
  \bibinfo {author} {\bibfnamefont {F.~J.}\ \bibnamefont
  {Rodr{\'\i}guez-Fortu{\~n}o}}, \ and\ \bibinfo {author} {\bibfnamefont
  {A.}~\bibnamefont {Mart{\'\i}nez}},\ }\href@noop {} {\bibfield  {journal}
  {\bibinfo  {journal} {Nanophotonics}\ }\textbf {\bibinfo {volume} {13}},\
  \bibinfo {pages} {4275} (\bibinfo {year} {2024})}\BibitemShut {NoStop}%
\bibitem [{\citenamefont {Mohammadi}\ and\ \citenamefont
  {Tagliabue}(2024)}]{mohammadi2024nanophotonic}%
  \BibitemOpen
  \bibfield  {author} {\bibinfo {author} {\bibfnamefont {E.}~\bibnamefont
  {Mohammadi}}\ and\ \bibinfo {author} {\bibfnamefont {G.}~\bibnamefont
  {Tagliabue}},\ }\href@noop {} {\bibfield  {journal} {\bibinfo  {journal} {ACS
  photonics}\ }\textbf {\bibinfo {volume} {12}},\ \bibinfo {pages} {152}
  (\bibinfo {year} {2024})}\BibitemShut {NoStop}%
\bibitem [{\citenamefont {Golat}\ \emph {et~al.}(2024)\citenamefont {Golat},
  \citenamefont {Kingsley-Smith}, \citenamefont {Diez}, \citenamefont
  {Martinez-Romeu}, \citenamefont {Mart{\'\i}nez},\ and\ \citenamefont
  {Rodr{\'\i}guez-Fortu{\~n}o}}]{golat2024optical}%
  \BibitemOpen
  \bibfield  {author} {\bibinfo {author} {\bibfnamefont {S.}~\bibnamefont
  {Golat}}, \bibinfo {author} {\bibfnamefont {J.~J.}\ \bibnamefont
  {Kingsley-Smith}}, \bibinfo {author} {\bibfnamefont {I.}~\bibnamefont
  {Diez}}, \bibinfo {author} {\bibfnamefont {J.}~\bibnamefont
  {Martinez-Romeu}}, \bibinfo {author} {\bibfnamefont {A.}~\bibnamefont
  {Mart{\'\i}nez}}, \ and\ \bibinfo {author} {\bibfnamefont {F.~J.}\
  \bibnamefont {Rodr{\'\i}guez-Fortu{\~n}o}},\ }\href@noop {} {\bibfield
  {journal} {\bibinfo  {journal} {Physical Review Research}\ }\textbf {\bibinfo
  {volume} {6}},\ \bibinfo {pages} {023079} (\bibinfo {year}
  {2024})}\BibitemShut {NoStop}%
\bibitem [{\citenamefont {Tetour}\ \emph {et~al.}(2024)\citenamefont {Tetour},
  \citenamefont {Novotn{\'a}}, \citenamefont {Tat{\`y}rek}, \citenamefont
  {M{\'a}kov{\'a}}, \citenamefont {Stuchl{\'\i}k}, \citenamefont {Hobbs},
  \citenamefont {{\v{R}}ezanka}, \citenamefont {M{\"u}llerov{\'a}},
  \citenamefont {Setni{\v{c}}ka}, \citenamefont {Dob{\v{s}}{\'\i}kov{\'a}}
  \emph {et~al.}}]{tetour2024preparations}%
  \BibitemOpen
  \bibfield  {author} {\bibinfo {author} {\bibfnamefont {D.}~\bibnamefont
  {Tetour}}, \bibinfo {author} {\bibfnamefont {M.}~\bibnamefont {Novotn{\'a}}},
  \bibinfo {author} {\bibfnamefont {J.}~\bibnamefont {Tat{\`y}rek}}, \bibinfo
  {author} {\bibfnamefont {V.}~\bibnamefont {M{\'a}kov{\'a}}}, \bibinfo
  {author} {\bibfnamefont {M.}~\bibnamefont {Stuchl{\'\i}k}}, \bibinfo {author}
  {\bibfnamefont {C.}~\bibnamefont {Hobbs}}, \bibinfo {author} {\bibfnamefont
  {M.}~\bibnamefont {{\v{R}}ezanka}}, \bibinfo {author} {\bibfnamefont
  {M.}~\bibnamefont {M{\"u}llerov{\'a}}}, \bibinfo {author} {\bibfnamefont
  {V.}~\bibnamefont {Setni{\v{c}}ka}}, \bibinfo {author} {\bibfnamefont
  {K.}~\bibnamefont {Dob{\v{s}}{\'\i}kov{\'a}}},  \emph {et~al.},\ }\href@noop
  {} {\bibfield  {journal} {\bibinfo  {journal} {Nanoscale}\ }\textbf {\bibinfo
  {volume} {16}},\ \bibinfo {pages} {6696} (\bibinfo {year}
  {2024})}\BibitemShut {NoStop}%
\bibitem [{\citenamefont {Onsager}(1936)}]{onsager1936electric}%
  \BibitemOpen
  \bibfield  {author} {\bibinfo {author} {\bibfnamefont {L.}~\bibnamefont
  {Onsager}},\ }\href@noop {} {\bibfield  {journal} {\bibinfo  {journal}
  {Journal of the American Chemical Society}\ }\textbf {\bibinfo {volume}
  {58}},\ \bibinfo {pages} {1486} (\bibinfo {year} {1936})}\BibitemShut
  {NoStop}%
\bibitem [{\citenamefont {Jackson}(1999)}]{jackson1999electrodynamics}%
  \BibitemOpen
  \bibfield  {author} {\bibinfo {author} {\bibfnamefont {J.~D.}\ \bibnamefont
  {Jackson}},\ }\href@noop {} {\emph {\bibinfo {title} {Classical
  Electrodynamics}}}\ (\bibinfo  {publisher} {John Wiley \& Sons, New York},\
  \bibinfo {year} {1999})\BibitemShut {NoStop}%
\bibitem [{\citenamefont {Stokes}(1851)}]{stokes1851composition}%
  \BibitemOpen
  \bibfield  {author} {\bibinfo {author} {\bibfnamefont {G.~G.}\ \bibnamefont
  {Stokes}},\ }\href@noop {} {\bibfield  {journal} {\bibinfo  {journal}
  {Transactions of the Cambridge Philosophical Society}\ }\textbf {\bibinfo
  {volume} {9}},\ \bibinfo {pages} {399} (\bibinfo {year} {1851})}\BibitemShut
  {NoStop}%
\bibitem [{\citenamefont {Crichton}\ and\ \citenamefont
  {Marston}(2000)}]{crichton2000measurable}%
  \BibitemOpen
  \bibfield  {author} {\bibinfo {author} {\bibfnamefont {J.~H.}\ \bibnamefont
  {Crichton}}\ and\ \bibinfo {author} {\bibfnamefont {P.~L.}\ \bibnamefont
  {Marston}},\ }\href@noop {} {\bibfield  {journal} {\bibinfo  {journal}
  {Electronic Journal of Differential Equations}\ }\textbf {\bibinfo {volume}
  {4}},\ \bibinfo {pages} {37} (\bibinfo {year} {2000})}\BibitemShut {NoStop}%
\bibitem [{\citenamefont {Olmos-Trigo}\ \emph {et~al.}(2024)\citenamefont
  {Olmos-Trigo}, \citenamefont {Nieto-Vesperinas},\ and\ \citenamefont
  {Molina-Terriza}}]{olmos2024spheres}%
  \BibitemOpen
  \bibfield  {author} {\bibinfo {author} {\bibfnamefont {J.}~\bibnamefont
  {Olmos-Trigo}}, \bibinfo {author} {\bibfnamefont {M.}~\bibnamefont
  {Nieto-Vesperinas}}, \ and\ \bibinfo {author} {\bibfnamefont
  {G.}~\bibnamefont {Molina-Terriza}},\ }\href@noop {} {\bibfield  {journal}
  {\bibinfo  {journal} {Physical Review Research}\ }\textbf {\bibinfo {volume}
  {6}},\ \bibinfo {pages} {043192} (\bibinfo {year} {2024})}\BibitemShut
  {NoStop}%
\bibitem [{\citenamefont {Hinamoto}\ \emph {et~al.}(2020)\citenamefont
  {Hinamoto}, \citenamefont {Hotta}, \citenamefont {Sugimoto},\ and\
  \citenamefont {Fujii}}]{hinamoto2020colloidal}%
  \BibitemOpen
  \bibfield  {author} {\bibinfo {author} {\bibfnamefont {T.}~\bibnamefont
  {Hinamoto}}, \bibinfo {author} {\bibfnamefont {S.}~\bibnamefont {Hotta}},
  \bibinfo {author} {\bibfnamefont {H.}~\bibnamefont {Sugimoto}}, \ and\
  \bibinfo {author} {\bibfnamefont {M.}~\bibnamefont {Fujii}},\ }\href@noop {}
  {\bibfield  {journal} {\bibinfo  {journal} {Nano Letters}\ }\textbf {\bibinfo
  {volume} {20}},\ \bibinfo {pages} {7737} (\bibinfo {year}
  {2020})}\BibitemShut {NoStop}%
\bibitem [{\citenamefont {Kerker}\ \emph {et~al.}(1983)\citenamefont {Kerker},
  \citenamefont {Wang},\ and\ \citenamefont
  {Giles}}]{kerker1983electromagnetic}%
  \BibitemOpen
  \bibfield  {author} {\bibinfo {author} {\bibfnamefont {M.}~\bibnamefont
  {Kerker}}, \bibinfo {author} {\bibfnamefont {D.-S.}\ \bibnamefont {Wang}}, \
  and\ \bibinfo {author} {\bibfnamefont {C.}~\bibnamefont {Giles}},\
  }\href@noop {} {\bibfield  {journal} {\bibinfo  {journal} {J. Opt. Soc. Am.
  A}\ }\textbf {\bibinfo {volume} {73}},\ \bibinfo {pages} {765} (\bibinfo
  {year} {1983})}\BibitemShut {NoStop}%
\bibitem [{\citenamefont {Olmos-Trigo}\ \emph {et~al.}(2020)\citenamefont
  {Olmos-Trigo}, \citenamefont {Abujetas}, \citenamefont {Sanz-Fern{\'a}ndez},
  \citenamefont {Zambrana-Puyalto}, \citenamefont {de~Sousa}, \citenamefont
  {S{\'a}nchez-Gil},\ and\ \citenamefont {S{\'a}enz}}]{olmos2020unveiling}%
  \BibitemOpen
  \bibfield  {author} {\bibinfo {author} {\bibfnamefont {J.}~\bibnamefont
  {Olmos-Trigo}}, \bibinfo {author} {\bibfnamefont {D.~R.}\ \bibnamefont
  {Abujetas}}, \bibinfo {author} {\bibfnamefont {C.}~\bibnamefont
  {Sanz-Fern{\'a}ndez}}, \bibinfo {author} {\bibfnamefont {X.}~\bibnamefont
  {Zambrana-Puyalto}}, \bibinfo {author} {\bibfnamefont {N.}~\bibnamefont
  {de~Sousa}}, \bibinfo {author} {\bibfnamefont {J.~A.}\ \bibnamefont
  {S{\'a}nchez-Gil}}, \ and\ \bibinfo {author} {\bibfnamefont {J.~J.}\
  \bibnamefont {S{\'a}enz}},\ }\href@noop {} {\bibfield  {journal} {\bibinfo
  {journal} {Physical Review Research}\ }\textbf {\bibinfo {volume} {2}},\
  \bibinfo {pages} {043021} (\bibinfo {year} {2020})}\BibitemShut {NoStop}%
\bibitem [{\citenamefont {Olmos-Trigo}\ and\ \citenamefont
  {Zambrana-Puyalto}(2022)}]{olmos2022helicity}%
  \BibitemOpen
  \bibfield  {author} {\bibinfo {author} {\bibfnamefont {J.}~\bibnamefont
  {Olmos-Trigo}}\ and\ \bibinfo {author} {\bibfnamefont {X.}~\bibnamefont
  {Zambrana-Puyalto}},\ }\href@noop {} {\bibfield  {journal} {\bibinfo
  {journal} {Physical Review Applied}\ }\textbf {\bibinfo {volume} {18}},\
  \bibinfo {pages} {044007} (\bibinfo {year} {2022})}\BibitemShut {NoStop}%
\bibitem [{\citenamefont
  {Olmos-Trigo}(2024{\natexlab{a}})}]{olmos2024revealing}%
  \BibitemOpen
  \bibfield  {author} {\bibinfo {author} {\bibfnamefont {J.}~\bibnamefont
  {Olmos-Trigo}},\ }\href@noop {} {\bibfield  {journal} {\bibinfo  {journal}
  {ACS photonics}\ }\textbf {\bibinfo {volume} {11}},\ \bibinfo {pages} {3697}
  (\bibinfo {year} {2024}{\natexlab{a}})}\BibitemShut {NoStop}%
\bibitem [{\citenamefont {Olmos-Trigo}(2024{\natexlab{b}})}]{olmos2024solving}%
  \BibitemOpen
  \bibfield  {author} {\bibinfo {author} {\bibfnamefont {J.}~\bibnamefont
  {Olmos-Trigo}},\ }\href@noop {} {\bibfield  {journal} {\bibinfo  {journal}
  {Nano letters}\ }\textbf {\bibinfo {volume} {24}},\ \bibinfo {pages} {8658}
  (\bibinfo {year} {2024}{\natexlab{b}})}\BibitemShut {NoStop}%
\bibitem [{\citenamefont {Mason}(1963)}]{Mason1963}%
  \BibitemOpen
  \bibfield  {author} {\bibinfo {author} {\bibfnamefont {S.~F.}\ \bibnamefont
  {Mason}},\ }\href {\doibase 10.1039/QR9631700020} {\bibfield  {journal}
  {\bibinfo  {journal} {Q. Rev. Chem. Soc.}\ }\textbf {\bibinfo {volume}
  {17}},\ \bibinfo {pages} {20} (\bibinfo {year} {1963})}\BibitemShut {NoStop}%
\bibitem [{Jas(2011)}]{Jasco2011}%
  \BibitemOpen
  \href@noop {} {\emph {\bibinfo {title} {Measurement of Vibrational Circular
  Dichroism spectra using the FVS-6000}}},\ \bibinfo {organization} {JASCO
  INTERNATIONAL CO., LTD.} (\bibinfo {year} {2011})\BibitemShut {NoStop}%
\bibitem [{\citenamefont {Salzman}\ and\ \citenamefont
  {Polavarapu}(1991)}]{Salzman1991}%
  \BibitemOpen
  \bibfield  {author} {\bibinfo {author} {\bibfnamefont {W.}~\bibnamefont
  {Salzman}}\ and\ \bibinfo {author} {\bibfnamefont {P.}~\bibnamefont
  {Polavarapu}},\ }\href {\doibase
  https://doi.org/10.1016/0009-2614(91)90282-E} {\bibfield  {journal} {\bibinfo
   {journal} {Chemical Physics Letters}\ }\textbf {\bibinfo {volume} {179}},\
  \bibinfo {pages} {1 } (\bibinfo {year} {1991})}\BibitemShut {NoStop}%
\bibitem [{\citenamefont {Barron}(2004)}]{Barron2004}%
  \BibitemOpen
  \bibfield  {author} {\bibinfo {author} {\bibfnamefont {L.~D.}\ \bibnamefont
  {Barron}},\ }\href@noop {} {\emph {\bibinfo {title} {Molecular Light
  Scattering and Optical Activity}}},\ \bibinfo {edition} {2nd}\ ed.\ (\bibinfo
   {publisher} {Cambridge University Press},\ \bibinfo {year}
  {2004})\BibitemShut {NoStop}%
\bibitem [{\citenamefont {Sihvola}\ \emph {et~al.}(1994)\citenamefont
  {Sihvola}, \citenamefont {Viitanen}, \citenamefont {Lindell},\ and\
  \citenamefont {Tretyakov}}]{Lindell1994}%
  \BibitemOpen
  \bibfield  {author} {\bibinfo {author} {\bibfnamefont {A.}~\bibnamefont
  {Sihvola}}, \bibinfo {author} {\bibfnamefont {A.}~\bibnamefont {Viitanen}},
  \bibinfo {author} {\bibfnamefont {I.}~\bibnamefont {Lindell}}, \ and\
  \bibinfo {author} {\bibfnamefont {S.}~\bibnamefont {Tretyakov}},\ }\href@noop
  {} {\emph {\bibinfo {title} {Electromagnetic Waves in Chiral and Bi-isotropic
  Media}}}\ (\bibinfo  {publisher} {Artech House},\ \bibinfo {year}
  {1994})\BibitemShut {NoStop}%
\bibitem [{\citenamefont {Vidal}\ \emph {et~al.}(2015)\citenamefont {Vidal},
  \citenamefont {Fernandez-Corbaton}, \citenamefont {Barbara},\ and\
  \citenamefont {Molina-Terriza}}]{Vidal2015}%
  \BibitemOpen
  \bibfield  {author} {\bibinfo {author} {\bibfnamefont {X.}~\bibnamefont
  {Vidal}}, \bibinfo {author} {\bibfnamefont {I.}~\bibnamefont
  {Fernandez-Corbaton}}, \bibinfo {author} {\bibfnamefont {A.~F.}\ \bibnamefont
  {Barbara}}, \ and\ \bibinfo {author} {\bibfnamefont {G.}~\bibnamefont
  {Molina-Terriza}},\ }\href {\doibase 10.1063/1.4936342} {\bibfield  {journal}
  {\bibinfo  {journal} {Appl. Phys. Lett.}\ }\textbf {\bibinfo {volume}
  {107}},\ \bibinfo {pages} {211107} (\bibinfo {year} {2015})}\BibitemShut
  {NoStop}%
\bibitem [{\citenamefont {Negoro}\ \emph {et~al.}(2023)\citenamefont {Negoro},
  \citenamefont {Sugimoto},\ and\ \citenamefont {Fujii}}]{negoro2023helicity}%
  \BibitemOpen
  \bibfield  {author} {\bibinfo {author} {\bibfnamefont {H.}~\bibnamefont
  {Negoro}}, \bibinfo {author} {\bibfnamefont {H.}~\bibnamefont {Sugimoto}}, \
  and\ \bibinfo {author} {\bibfnamefont {M.}~\bibnamefont {Fujii}},\
  }\href@noop {} {\bibfield  {journal} {\bibinfo  {journal} {Nano Letters}\ }
  (\bibinfo {year} {2023})}\BibitemShut {NoStop}%
\bibitem [{\citenamefont {Sugimoto}\ \emph {et~al.}(2020)\citenamefont
  {Sugimoto}, \citenamefont {Okazaki},\ and\ \citenamefont
  {Fujii}}]{sugimoto2020mie}%
  \BibitemOpen
  \bibfield  {author} {\bibinfo {author} {\bibfnamefont {H.}~\bibnamefont
  {Sugimoto}}, \bibinfo {author} {\bibfnamefont {T.}~\bibnamefont {Okazaki}}, \
  and\ \bibinfo {author} {\bibfnamefont {M.}~\bibnamefont {Fujii}},\
  }\href@noop {} {\bibfield  {journal} {\bibinfo  {journal} {Advanced Optical
  Materials}\ }\textbf {\bibinfo {volume} {8}},\ \bibinfo {pages} {2000033}
  (\bibinfo {year} {2020})}\BibitemShut {NoStop}%
\end{thebibliography}%

\clearpage

\appendix 
\onecolumngrid

\section{The Stokes vector method for spherical chiral nanoparticles} \label{A_1}

In this Section, we derive Eq.~(6) of the main text. 
After algebraic manipulation of
Eq.~(3) we arrive to
\begin{equation} \label{e_fields_1}
\E_{\rm{sca}}(k \r) = {k^2} \frac{e^{ikr}}{4\pi r} \left[ \left( \frac{p_\theta}{\epsilon}+ Z m_ \varphi \right) \hat{\mathbf{e}}_{\rm{\theta}} + \left( \frac{p_\varphi}{\epsilon} - Z m_ \theta \right) \hat{\mathbf{e}}_{\rm{\varphi}}  \right],
\end{equation}
where $\hat{\mathbf{e}}_\theta$ and $\hat{\mathbf{e}}_\varphi$ are unit vectors in spherical coordinates.
We now express the incident electromagnetic field in spherical coordinates. In particular, we consider the simplest incident wavefield carrying well-defined helicity $\sigma = \pm 1$: a circularly polarized plane wave propagating in the $z$-direction.  The electric field of such plane wave evaluated at $z = 0$ is given by
\begin{equation} \label{pw}
\frac{\E_{\rm{inc}}}{E_0} = \frac{ \left(\hat{\mathbf{e}}_x + i \sigma \hat{\mathbf{e}}_y \right)}{\sqrt{2}}  =  e^{i \sigma \varphi }\left[ \frac{ \sin \theta \hat{\mathbf{e}}_r + \cos \theta \hat{\mathbf{e}}_\theta + i \sigma \hat{\mathbf{e}}_\varphi }{\sqrt{2}} \right],
\end{equation}
where $E_0$ is the amplitude of the incident plane-wave.
The electric and magnetic dipoles can be compactly written as 
\begin{align} \label{p_exp}
\frac{\p}{\epsilon} &=   \gamma^\sigma \E_{\rm{inc}},  & \gamma^\sigma &=\alpha_{\rm{ee}} - i \sigma \alpha_{\rm{em}}, \\ 
\label{m_exp}
Z \m &= -i \sigma \beta^\sigma \E_{\rm{inc}},  &
\beta^\sigma &=   \alpha_{\rm{mm}} -i \sigma \alpha_{\rm{em}}  .
\end{align}
Now, expanding  Eqs.~\eqref{p_exp}-\eqref{m_exp} using Eq.~\eqref{pw} we have 
\begin{equation} \label{dip_1}
\frac{p_\theta}{\epsilon} + Z m_\varphi = E_0 e^{i \sigma \varphi } \left( \frac{ \gamma^\sigma \cos \theta +  \beta^\sigma }{\sqrt{2}} \right),  
\end{equation}
\begin{equation} \label{dip_2}
\frac{p_\varphi}{\epsilon} - Z m_\theta = i \sigma E_0 e^{i \sigma \varphi }  \left( \frac{ \gamma^\sigma +\beta^\sigma \cos \theta }{\sqrt{2}} \right).  
\end{equation}
Inserting Eqs.~\eqref{dip_1}-\eqref{dip_2} into Eq.~\eqref{e_fields_1} we get $
\E_{\rm{sca}} = E_{\rm{\theta}}\hat{\mathbf{e}}_{\rm{\theta}} \ + {E}_{\rm{\varphi}}\hat{\mathbf{e}}_{\rm{\varphi}}$, where
\be \label{ddd}
{E_{\rm{\theta}}} &=&C(kr, \varphi) \left[ \gamma^\sigma \cos \theta +  \beta^\sigma\right], \\ \label{e_phdddi}
{{E}_{\rm{\varphi}}} &=& i \sigma C(kr, \varphi)  \left[ \gamma^\sigma + \beta^\sigma \cos \theta \right].
\ee 
Here 
$C(kr, \varphi) =  E_0 \sqrt{2} k^2{e^{i(kr+ \sigma \varphi)}}/{(8 \pi r)}$.  We now insert  Eq.~\eqref{ddd}-\eqref{e_phdddi} into the Stokes parameters. After some algebra, we have 
\begin{align} \label{s0_new}
{s}^\sigma_0 (kr, \theta) &= |C(kr, \varphi)|^2 \left[(|\gamma^\sigma|^2 + |\beta^\sigma|^2) \left(\cos^2 \theta + 1 \right)  + 4 \Re \{\gamma^\sigma ({\beta^\sigma})^* \} \cos \theta \right], \\
{s}^\sigma_1 (kr, \theta) &= |C(kr, \varphi)|^2 (|\gamma^\sigma|^2 - |\beta^\sigma|^2) \left(\cos^2 \theta -1 \right), \\ \label{mystery}
{s}^\sigma_2 (kr, \theta) &= -2 \sigma |C(kr, \varphi)|^2  \Im \{\gamma^\sigma (\beta^\sigma)^* \} \left(\cos^2 \theta -1 \right), \\ \label{s3_new}
{s}^\sigma_3 (kr, \theta) &= - 2 \sigma |C(kr, \varphi)|^2  \left( (|\gamma^\sigma|^2 + |\beta^\sigma|^2)\cos \theta  + \Re \{\gamma^\sigma ({\beta^\sigma})^* \} \left(\cos^2 \theta  + 1 \right) \right). 
\end{align}
These equations can be conveniently written in a matrix representation. That is, 
  \begin{equation} \label{mid_way}
   \begin{pmatrix}
    {s}^\sigma_0 (kr, \theta)\\
    {s}^\sigma_1 (kr, \theta)\\ {s}^\sigma_2 (kr, \theta)\\
    {s}^\sigma_3 (kr, \theta)
\end{pmatrix} = |C(kr, \varphi)|^2
    \begin{pmatrix}
    1 + \cos^2 \theta & 1 + \cos^2 \theta & 4 \cos \theta & 0 \\
   -\sin^2 \theta  & \sin^2 \theta& 0 & 0  \\
    0 & 0 & 0 &  2\sigma \sin^2 \theta \\
    -2\sigma \cos \theta  &   -2\sigma \cos \theta  & -2 \sigma \left(1 + \cos^2 \theta\right) & 0  
\end{pmatrix}
 \begin{pmatrix}
 |\gamma^\sigma|^2 \\
 |\beta^\sigma|^2\\ \Re \{\gamma^\sigma (\beta^\sigma)^* \}\\
    \Im \{\gamma^\sigma (\beta^\sigma)^* \}
    \end{pmatrix}.
\end{equation}  
 Inverting Eq.~\eqref{mid_way}, we arrive to
  \begin{equation} \label{compact}
 \underbrace{\begin{pmatrix}
 |\gamma^\sigma|^2 \\
 |\beta^\sigma|^2\\ \Re \{\gamma^\sigma (\beta^\sigma)^* \}\\
    \Im \{\gamma^\sigma (\beta^\sigma)^* \}
    \end{pmatrix}}_{ \let\scriptstyle\textstyle\substack{\mathbf{J}^\sigma}} = \underbrace{\frac{32 \pi ^2 (kr)^2 }{k^6 |E_0|^2}  \frac{\csc^4 \theta}{2}
    \begin{pmatrix}
    1 + \cos^2 \theta & - \sin^2 \theta& 0 & 2 \sigma \cos \theta \\
    1 + \cos^2 \theta  & \sin^2 \theta& 0 &  2 \sigma \cos \theta  \\
    -2 \cos \theta & 0 & 0 & -\sigma \left(1 + \cos^2 \theta \right)\\
    0 & 0 & \sigma \sin^2 \theta & 0  
\end{pmatrix}}_{\let\scriptstyle\textstyle\substack{{\til{U}^\sigma (kr, \theta)}}}
\underbrace{
 \begin{pmatrix}
    {s}^\sigma_0 (kr, \theta)\\
    {s}^\sigma_1 (kr, \theta)\\ {s}^\sigma_2 (kr, \theta)\\
    {s}^\sigma_3 (kr, \theta)
\end{pmatrix}}_{\let\scriptstyle\textstyle\substack{{{\mathbf{S}}^\sigma(kr, \theta)}}},
\end{equation}
which is identical to Eq.~(6) of the main text.

\end{document}